\documentclass[prb,reprint,nofootinbib]{revtex4-1}
\bibpunct{[}{]}{;}{n}{}{}

\pdfoutput=1
\usepackage[caption=false]{subfig}

\usepackage{graphicx}
\usepackage{comment}
\usepackage{amsmath}
\usepackage{amssymb}
\usepackage{amsfonts}
\usepackage{bbm}
\usepackage{braket}
\usepackage{verbatim}
\usepackage[hidelinks]{hyperref}
\usepackage{ulem}
\hypersetup{
  colorlinks   = true, 
  urlcolor     = blue, 
  linkcolor    = blue, 
  citecolor   = blue 
}

\usepackage{color}
\usepackage{tikz}

\newcommand{\be}{\begin{equation}}
\newcommand{\ee}{\end{equation}}

\newcommand{\tr}[1]{\text{Tr}\big[{#1}\big]}

\begin{document}
	
	\title{Emergent conservation laws and nonthermal states in the mixed-field Ising model}
	
	\author{Jonathan Wurtz}
	\email[Corresponding author: ] {jwurtz@bu.edu}
	\affiliation{Department of Physics, Boston University, 590 Commonwealth Ave., Boston, MA 02215, USA}

	\author{Anatoli Polkovnikov}
	\affiliation{Department of Physics, Boston University, 590 Commonwealth Ave., Boston, MA 02215, USA}

	\begin{abstract}

    This paper presents a method of computing approximate conservation laws and eigenstates of integrability-broken models using the concept of adiabatic continuation. Given some Hamiltonian, eigenstates and conserved operators may be computed by using those of a simple Hamiltonian close by in parameter space, dressed by some unitary rotation. However, most adiabatic continuation analyses only use this unitary implicitly. In this work, approximate adiabatic gauge potentials are used to construct a state dressing using variational methods, to compute eigenstates via a rotated truncated spectrum approximation. These methods allow construction of both low and high-energy approximate nonthermal eigenstates, as well as quasi-local almost-conserved operators, in models where integrability may be non-perturbatively broken. These concepts will be demonstrated in the mixed-field Ising model.
	\end{abstract}

	\maketitle


    Adiabatic continuation is a well-used concept \cite{Hastings2005}. Given some Hamiltonian of interest, low energy eigenstates of that system may be computed by finding a ``simple" Hamiltonian ``nearby" in parameter space, which can be easily diagonalized. Then, those simple eigenstates can be mapped to interacting ones using some dressing by a unitary $U$. This procedure can be used to show that ground states can be mapped to other ground states, as long as there is some path in parameter space which does not go through some gap-closing critical point \cite{Nachtergaele2006,hastings2007,Xie2010}. For example, low energies of interacting fermions may be described by a Fermi gas or Fermi liquid with dressed quasiparticle excitations \cite{Shankar1994}. However, most of these proofs are non-constructive in the sense that the unitary $U$ is generally never actually computed, or only done so perturbatively.
    
    This work details explicitly connecting non-interacting trivial states to nontrivial interacting ones, by variationally computing an approximate, local adiabatic gauge potential (AGP) to construct the unitary transformation. The latter modifies non-interacting particle states to quasiparticle states which are ``dressed" within some local span of sites. Importantly, the dressing is non-perturbative and not limited to low-energy states. This construction leads to long-lasting quasiparticles and stable nonthermal states, even at finite energy densities.
    
    As a consequence of computing unitary rotations for approximate eigenstates, this procedure also allows one to compute local almost-conserved operators from the approximate eigenstates and dressed non-interacting symmetries. In the presence of integrability breaking terms in some system, undressed symmetries and conserved operators are generally no longer conserved~\cite{Langen2016,Bertini2016,Yijun2018}: instead, these operators may be ``dressed" locally by the unitary $U$ to restore approximate conservation of some quasi-local and long lived operator, even though the full system may no longer be integrable \cite{serbyn2013,Inbrie2016,Abanin2017}. Similarly, for particular initial wavefunctions with large overlap with good approximate eigenstates, one may compute long time effective quench dynamics within a small subspace of states, in spirit of Schrieffer-Wolff transformations \cite{Wurtz2020}.
    
	The presence of such unitary dressings, quasi-local conservation laws, and good approximate eigenstates in interacting models may suggest that not all integrability-broken models should be treated equally. Certain models may be ``close to integrable", in the sense that there exists a good local dressing of particular eigenstates, and strong ETH may be violated. Two particular cases of such ETH violation are many body localization \cite{Abanin2019} and quantum scars in the PXP model \cite{Bernien2017,turner2018b}. Conversely, other models may be far from any simple system, in the sense that there is no path in parameter space that admits a good local adiabatic dressing, and the model is quantum chaotic \cite{Srednicki1994}. In fact, this unitary rotation may potentially be seen as the analog of the canonical transformation of KAM theory which restores integrability in classical models \cite{Brandino2015}. While this paper focuses on a quantum model, the whole methodology, including variationally computed canonical transformations generated by the AGP~\cite{Polkovnikov2017}, is fully applicable to classical non-integrable systems.
    
    The rest of this paper is structured as follows. First will describe the methods of computing approximate eigenstates with the variational adiabatic gauge potential and the rotated truncated spectrum approach (rTSA). The method is a different perspective of Schrieffer-Wolff block diagonalization methods, as discussed in Ref. \cite{Wurtz2020}. Next will introduce the specific model used in this work, the non-integrable mixed-field Ising spin 1/2 chain, and demonstrate the performance of computing approximate eigenstates. Finally will be an example of an approximately conserved local operator for the mixed-field Ising chain, the total quasiparticle number.

    \section{Computing local approximate eigenstates}\label{sec:methods}
    
    In general, directly computing eigenstates and quasiparticle excitations is hard, as computational difficulty scales exponentially in system size, and normally there are no well-defined quantum numbers. Instead, let us continue in spirit of adiabatic continuation \cite{Hastings2005}. Suppose some parameterized Hamiltonian $H(\mu)$, with $\mu=1$ being the particular system of interest, and $\mu=0$ being exactly solvable, in the sense that the eigenstates are easily computable via symmetry, integrability, or other means. $\mu$ describes some choice of path in a (multi)-parameter space of Hamiltonians between 0 and 1.

    \begin{equation}
        \begin{tabular}{c c c c c}
             \\$H(0)$&$\Longrightarrow H(\mu)\Longrightarrow$& $H(1)$\\
             &\text{Path from 0 to 1}&\\
             \text{Exactly Solvable} &$\Downarrow$&\text{System of Interest}\\
             &\text{AGP}&\\
             $|E_n(0)\rangle$&$\Longrightarrow \mathcal A(\mu)\Longrightarrow$&$|E_n(1)\rangle$
        \end{tabular}\nonumber
    \end{equation}
    
    At all points along this path, there are parameterized eigenstates $\{|E_n(\mu)\rangle\}$ with eigenenergies $\{E_n(\mu)\}$. How these eigenstates change as a function of parameter $\mu$ is given by the adiabatic gauge potential (AGP) $\mathcal A(\mu)$ \cite{Polkovnikov2017}, computed from the instantaneous Hamiltonian $H(\mu)$

    \begin{equation}\label{eq:exact_evolve}
        i\partial_\mu |E_n(\mu)\rangle = \mathcal A(\mu)|E_n(\mu)\rangle.
    \end{equation}
    
    In principle, given the (parameter dependent) AGP $\mathcal A(\mu)$, one could use Eq. \eqref{eq:exact_evolve} to evolve the simple eigenstates $|E_n(0)\rangle$ into exact interacting eigenstates $|E_n(1)\rangle$, or equivalently use a unitary rotation
    
    \begin{align}\label{eq:unitary_rotation}
        U^\dagger =& \mathcal T \exp\bigg(i\int_0^1 \mathcal A(\mu)d\mu\bigg)
    \end{align}
    such that $|E_n(1)\rangle = U|E_n(0)\rangle$, with $\mathcal T$ indicating path ordering. In practice, this recipe runs into the same infeasibility as directly computing eigenstates: computing the exact AGP is generally as difficult as computing the exact eigensystem. It is nonlocal, exponentially large, and highly parameter dependent \cite{Polkovnikov2017} for generic interacting systems, making computing the unitary a similarly difficult task.
    
    \quad
    
    To circumvent the problems of computing the exact AGP, one can instead use \emph{approximate} adiabatic gauge potentials. One might hope that simple eigenstates ``dressed" by a local approximation of the AGP will closely resemble eigenstates of the full system. As the complexity of the approximation grows, so too should the approximate AGP approach the exact one, and the approximate eigenstates become exact, at the expense of them being highly entangled and nonlocal.
    
    From an adiabatic continuation standpoint, the connection with this approximate gauge potential is clear; in fact M. Hastings in \cite{Hastings2005} has is a particular implementation of an approximate AGP for gapped ground states. A simplified version of  Eq. (17) in Ref. \cite{Hastings2005} is equivalent to Eq.~\eqref{eq:unitary_rotation} with
    
        \begin{align}
        \mathcal A(\mu)\approx -\frac{1}{2}\int_{-\infty}^\infty  \text{SGN}[t] f(t)[\partial_\mu H](t)dt,\label{eq:regularized_AGP}\\
        f(t) = \text{erfc}\bigg(\bigg|\frac{t}{\tau_q}\bigg|\bigg)
    \end{align}
    where $\partial_\mu H(t)$ is the operator $\partial_\mu H$ in the Heisenberg representation, erfc is the complementary error function for which $\text{erfc}(0)=1$ and erfc$(\infty)=0$, and SGN$(t)$ is $\pm1$ depending on sign of $t$ (see Appendix \ref{app:hastings05} for more details). This expression 
    is nothing but an approximation of the gauge potential with a particular choice of regularizer $f(t)$ \cite{Claeys2019}. For the regularization time $\tau_q\to\infty$, this approximate AGP becomes exact. For a finite regularization time, the AGP is approximate but local within some span of sites.
    
    Instead of computing an approximate gauge potential via some choice of regulator, which is computationally difficult, one can instead compute it variationally \cite{Sels2017}. Here, some ansatz for the variational gauge potential is chosen based on the system at hand, then variationally optimized to best approximate the exact AGP. In this case the variational AGP is chosen to be the sum of some set of (local) operators $\{B_i\}$ for variational parameters $\{\alpha_i\}$
    
    \begin{equation}
    A(\{\alpha\}) = \sum_i \alpha_i B_i.
    \end{equation}
    
    The particular choice of operators $\{B_i\}$ depends on the problem at hand; if the set is expanded to include all operators in the space modulo symmetries (a potentially exponential number), the exact AGP can be reconstructed from the set, and the variational version becomes exact. Thus, one should expect improvement as the size of the ansatz is expanded. It has been found \cite{Claeys2019} that a local variational AGP can accurately reproduce the exact one, at least for matrix elements which are far separated in energy or protected by approximate symmetries. This is shown by using a Taylor series expansion of Eq. \eqref{eq:regularized_AGP}. In fact, this argument is mirrored by an equivalent justification from an adiabatic continuation perspective for gapped ground states: As long as there is a gap, the ground state will only entangle within some light cone \cite{lieb1972,Nachtergaele2006,hastings2007} and thus a local approximation for the rotation generator is always possible.
    
    The optimization of the AGP is computed by finding the minimum of \cite{Polkovnikov2017,Sels2017}
    
    \begin{equation}\label{eq:VarOptimize}
        S(\alpha)=\bigg|\bigg|\big[H,\partial_\mu H + i[A(\alpha),H]\big]\bigg|\bigg|
    \end{equation}
    where parameter dependence of $\alpha$ and $H$ on $\mu$ is implicit, and $||Q||=\tr{Q^2}/\mathcal D$ is the Hilbert-Schmidt norm. Eq. \eqref{eq:VarOptimize} has the property that $S=0$ for the exact AGP, and is solvable even for very large system sizes. This is because the function $S(\alpha)$ can be computed using trace identities, and is quadratic in $\alpha$, making optimization simple. This minimization leads to a quasi-local operator $A(\mu)$ which approximates the exact AGP.
    
    \quad
    
    One can then use this approximate AGP to compute approximate eigenstates, in the same manner of the exact AGP computing exact eigenstates. First, choose some set of $\mathcal D_p$ states $\{|q\rangle\}$ of the exactly solvable Hamiltonian $H(0)$. This choice depends on the system at hand. One choice could be, for example, all eigenstates below some energy cutoff. Another choice would be all states within some particular symmetry sector such as fixed particle number, which form a subset of eigenstates not necessarily sorted by energy. The set of states forms some projective subspace $P$ with projector $\mathcal P = \sum_q |q\rangle\langle q|$.
    
    Approximate eigenstates of the interacting model $H(1)$ can be computed by ``dressing" each state via Schr\"odinger evolution of Eq. \eqref{eq:exact_evolve} to implement the unitary of Eq. \eqref{eq:unitary_rotation}. This gives some set of ``dressed" states $\{|q(1)\rangle\}\equiv \{U|q\rangle\}$.
    
    As a comment on implementation, care must be taken in the direction of evolution: the AGP is parameter dependent so generally $A(1)\neq A(0)$. For perturbative couplings this parameter dependence is very weak so that the AGP approximately commutes with itself for all $\mu$ and the directionality doesn't matter; however strong coupling may lead to nonsensical answers. One may start by accidentally acting on a non-interacting wavefunction with the AGP from the interacting point, which may be much different than the correct non-interacting AGP.
    
    An effective Hamiltonian within that subspace can be computed via matrix elements in the dressed subspace:
    
    \begin{equation}
        \big(H_\text{eff}\big)^{pq} = \langle p(1)| H(1) |q(1)\rangle=\langle p|U^\dagger H(1) U|q\rangle.
    \end{equation}
    
    Note that this is equivalent to computing the effective Schrieffer-Wolff Hamiltonian \cite{Wurtz2020}, where one rotates the operator $\tilde H = U^\dagger H(1) U$ instead of the states. If the AGP is exact, this effective Hamiltonian will be exactly diagonal. However, if the AGP is not exact or the initial subspace was a degenerate symmetry sector, the effective Hamiltonian will not be diagonal. One can then compute the eigensystem of the $\mathcal D_p\times \mathcal D_p$ matrix via standard linear algebra techniques to find eigenenergies $E_i$ and eigenvectors $V_i$ such that
    
    \begin{equation}\label{eq:re-diagonalization1}
    \big(H_\text{eff}\big)^{nm} V_{mi} = E_i V_{ni}.
    \end{equation}
    
    The approximate eigenvectors of the system are then given by
    \begin{equation}\label{eq:re-diagonalization2}
        |E_i\rangle = \sum_q V_{qi}|q(1)\rangle
    \end{equation}
    with eigenvalues $E_i$. This final step is functionally equivalent to the truncated spectrum Approach (TSA) \cite{James2017}, except instead of using a non-interacting subspace $P$, the subspace is first rotated by the approximate AGP to obtain some improved subspace $\tilde P$. This basis better resembles eigenvectors of the interacting system, and may be exponentially orthogonal from the original basis due to the finite rotation. This corresponds to a \emph{rotated} truncated spectrum approach (rTSA)

    \begin{table} 
    \rule{\linewidth}{0.6pt}
    
    The abbreviated method is as follows:
    
    \begin{enumerate}
        \item Define some Hamiltonian $H(\mu)$, with $H(0)$ being exactly solvable and $H(1)$ being a system of interest, with some path in parameter space linking the two.
        \item Given some ansatz, compute a variational adiabatic gauge potential $A(\mu)$ along the points $\mu\in[0,1]$.
        \item Define some set of eigenstates of $H(0)$, either within some energy window or within some symmetry sector(s) such as particle number.
        \item Evolve the set of states via the Schr\"odinger equation from $\mu=0$ to $\mu=1$ with the variational AGP.\label{item:4}
        \item Compute the effective Hamiltonian and its eigensystem to find approximate eigenstates and eigenvalues.
    \end{enumerate}
    \rule{\linewidth}{0.6pt}
    \end{table}
    
    Because the variational AGP is local by construction, various tricks can be employed to go to large or even thermodynamic system sizes. As commented above, computing the variational AGP is not a problem for a large number of sites, as complexity scales linearly with system size. It is reasonable to compute an AGP for hundreds of sites for all operators spanning less than 5-6 sites on a modern desktop. This locality can also be used for step \ref{item:4} when evolving the basis states: due to the finite evolution ``time", states are only entangled within some finite region (using intuition of Lieb-Robinson bounds \cite{hastings2007}). This suppresses the finite size effects of evolving some small (typically 15-20) number of sites exactly, and enables tensor methods such as matrix product states (MPS) and variational evolution \cite{Verstraete2008}. This work employs the former method of exact evolution on small systems \cite{Weinberg2017}.
    
    Choosing a larger subspace should also be expected to improve the computation of the approximate eigensystem. In the limit where the projective subspace is the full Hilbert space or within one of the symmetries of the full Hamiltonian, the effective Hamiltonian is the exact one and likewise the eigenstates are exact, independent of the rotation. Choosing a subspace within some larger energy window should also be expected to improve the variational dressing: the AGP fails to suppress excitations close together in energy, but those can then be recaptured within the subspace $P$ via off-diagonal elements of the effective Hamiltonian.
    
    Because both the variational AGP ansatz and the projective subspace can be systematically expanded, this method gives a controllable approximation to compute eigenstates: as the complexity increases, the eigenstates will asymptotically approach the exact ones.
    
    The eigenstates computed in this manner are approximate, in that they are not exact eigenstates of the Hamiltonian $H(1)$. The simplest indicator of the closeness to an exact eigenstate is the energy variance of the state
    
    \begin{equation}\label{eq:energyvariance}
        \Delta_n^2 \equiv \langle E_n|H^2|E_n\rangle - \big|\langle E_n|H|E_n\rangle\big|^2.
    \end{equation}
    
    Exact eigenstates have zero energy variance, and so approximate eigenstates should have minimal energy variance $\Delta_n^2\approx 0$. The average energy variance of these eigenstates within the subblock corresponds to the average block-off-diagonal matrix elements in the Hamiltonian and thus indicates the performance of the block diagonalization procedure (see Appendix \ref{app:Eflucs}).
    
    \section{Model}

    As a concrete example, suppose the following system, the mixed field Ising model
    
    \begin{equation}\label{eq:base_hamiltonian}
        H = \sum_i^NJ\sigma_z^i\sigma_z^{i+1} + h_x\sigma_x^i + h_z\sigma_z^i.
    \end{equation}
    
    For $h_z=0$ the model is integrable via a Jordan-Wigner transformation to free fermions \cite{Sachdev2011a,Calabrese2012a,Calabrese2012b} with a critical point at $h_x=1$ and small $h_z$ being an integrable E(8) field theory \cite{zamolodchikov2989}. For $h_x=0$ the model is a purely classical Ising model. For $h_x^2+h_z^2\to\infty$ the model is an exactly solvable collection of single spins with an onsite field. At $h_z=2,h_x=0$ there is a first-order multicritical point \cite{Simon2011} and for small $h_x$, the low energy effective Hamiltonian is the PXP model \cite{Bernien2017,Turner2018a}. Elsewhere, the model has no apparent conservation laws or symmetries beyond geometric ones and is generally quantum chaotic \cite{Huse2014,dalessio2016}. However, this does not prevent approximate conservation laws or nonthermal states, as will indeed be seen.
    
    The variational ansatz is chosen to be that of Jordan-Wigner strings, i.e. strings of Pauli operators which map to fermion bilinear operators, plus all operators local within a span of $n$ sites
    
    \begin{align}\label{eq:AGPansatz}
        \{B\}=\{&\sigma_y^0,\quad \sigma_x^0\sigma_y^1,\dots,\sigma_x^0\sigma_y^n,\dots,\sigma_y^0\sigma_z^1\sigma_x^2,\dots,\\
        &\sigma_z^0\sigma_y^1,\quad\sigma_z^0\sigma_x^1\sigma_y^2,\quad\sigma_z^0\sigma_x^1\sigma_x^2\sigma_y^3,\quad \sigma_z^0\sigma_x^1\sigma_x^2\sigma_x^3\sigma_y^4\dots\nonumber
        \}.
    \end{align}
    
    Additional symmetries and properties reduce the size of the ansatz: the AGP has all of the symmetries of the full Hamiltonian \footnote{This can be seen by the definition of the regularized AGP in Eq. \eqref{eq:regularized_AGP}}. By gauge choice the AGP can be completely imaginary for real Hamiltonians \cite{Polkovnikov2017}, constraining $\{B\}$ to only include terms with an odd number of $\sigma_y$. Because the Hamiltonian is translation and reflection invariant, the ansatz can be chosen to be as well. The inclusion of Jordan-Wigner strings is motivated by this ansatz being exact for the transverse Ising model \cite{Polkovnikov2017}, due to its extra symmetries and mapping to free fermions.

    While a Hamiltonian of interest is given by particular choice of parameters $h_x$, $h_z$, there is relative freedom for choice of the path $h_x(\mu)$, $h_z(\mu)$ in the 2d parameter space $H(\mu)$, and especially choice of simple Hamiltonian $H(0)$. This is because there are many ``simple" points in the $(h_x,h_z)$ parameter space which might be considered ``close" to the Hamiltonian of interest. The $(h_x,0)$ line is the transverse Ising model; the $(0,h_z)$ line is the classical Ising model; and the $(h_x,h_z)\to\infty$ line are independent spins with onsite fields.
    
    What starting points, and which path in parameter space, is optimal for computing approximate eigenstates, given ansatz $\{B\}$, Hamiltonian $H(1)$, and subspace P? This is a question of a \emph{path-dependent Schrieffer Wolff transformation}, as the performance of computing approximate eigenstates, or equivalently block diagonalization, may depend on these choices. This work chooses from a limited set of parameterized Hamiltonians with particular starting and ending points. Two additional parameterizations are discussed in Appendix \ref{app:pathdependence}.
    
    \begin{align}
        &H_1(\mu) = \sum_i^N \sigma_z^i\sigma_z^{i+1} + \mu\big(h_x \sigma_x^i + h_z\sigma_z^i\big),\label{eq:H1}\\
        &H_2(\mu) = \sum_i^N -\sigma_z^i\sigma_z^{i+1} + \mu\big(h_x \sigma_x^i + h_z\sigma_z^i\big),\label{eq:H2}\\
        &H_3(\mu) = \begin{cases}
        \sum_i^N 2\mu h_x \sigma_x^i + h_z\sigma_z^i,&\mu\in[0,0.5)\\\\
        \sum_i^N (2\mu-1)\sigma_z^i\sigma_z^{i+1}&\mu\in[0.5,1]\\
        \quad\quad  + \big(h_x \sigma_x^i + h_z\sigma_z^i\big)
        \end{cases}.\label{eq:H3}
    \end{align}

\begin{figure}
        \centering
        \includegraphics[width=\linewidth]{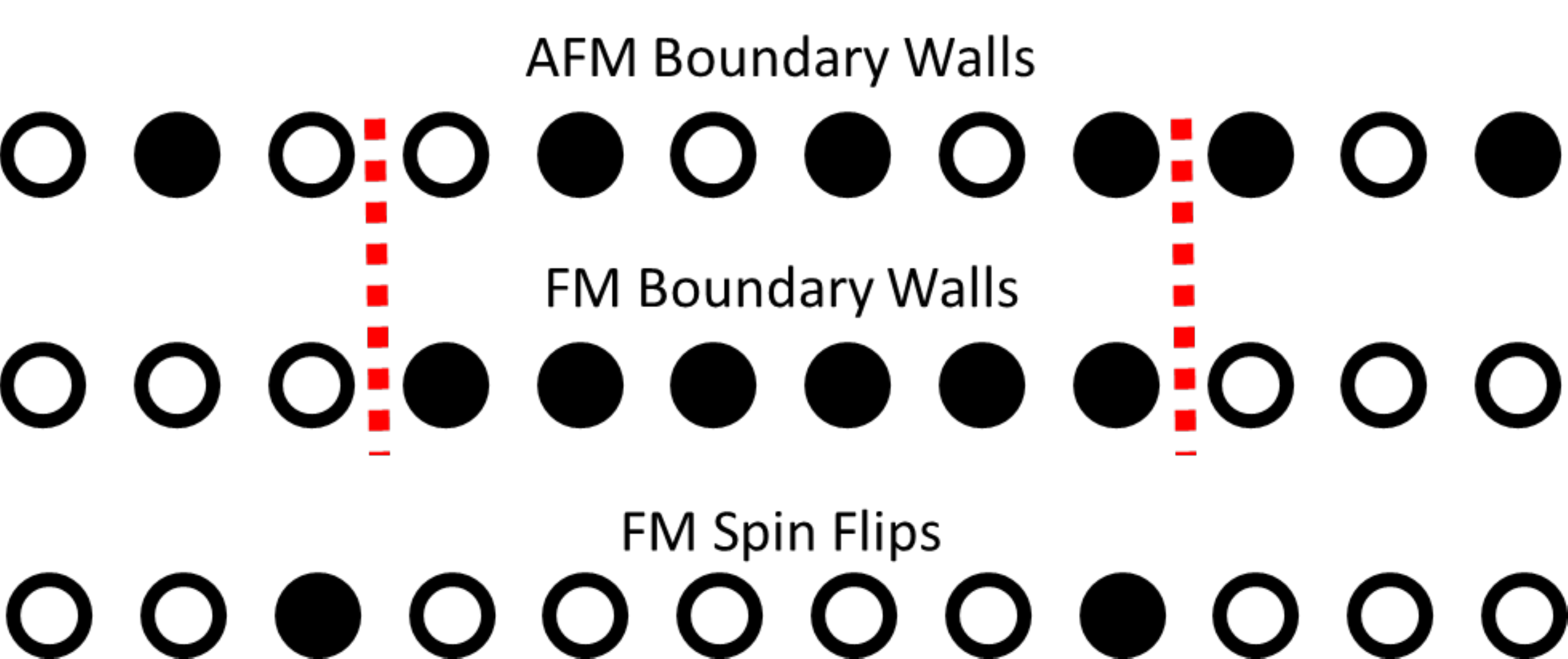}
        \caption{Example basis states of two particles separated by 6 sites. Top and middle are states with two boundary walls (red dashes), which are low energy eigenstates of $H_1(0)$ and $H_2(0)$, with excitation energy $4J$. Bottom is a state with two spin flip particles, which is a low-energy state of $H_3(0)$ with excitation energy $4h_z$.}
        \label{fig:state_examples}
    \end{figure}

    The first and second parameterizations start from the $\sigma_z\sigma_z$ point, whose eigenstates are $Z$ polarized spins. Depending on the sign, the ground state could be an antiferromagnetic (AFM) Ne\'el \eqref{eq:H1} or a polarized ferromagnetic (FM) state \eqref{eq:H2}. Low energy particle excitations are boundary walls of spin flips (see Fig. \ref{fig:state_examples}) \footnote{This work exclusively uses an even number of sites and periodic boundary conditions to avoid any AFM ground state degeneracy; particles always come in pairs.}.
    
    The third parameterization \eqref{eq:H3} is split into two parts. The first leg is simply rotating the on-site field and thus the AGP is exact $A(\mu)\sim \sum_i\sigma_y$, and is an example of the Landau-Zener problem, rotating the spin in the XZ plane. The second leg has no such local exact representation. The ground state is a product state of spins pointing in Z. Low energy particle excitations are spin flips (see Fig. \ref{fig:state_examples}) to the opposite direction.
    
    In all cases, $H_*(0)$ is degenerate, with a natural choice of projective subspace being fixed particle number on top of the ground state. Thus, $P_1$ is 0 and 2 boundary walls on top of an AFM ground state; $P_2$ is 0 and 2 boundary walls on top of a FM state; and $P_3$ is the 0, 1 and 2 particle spin flips on top of a polarized state.
    
    \begin{align}\label{eq:subspace_defns}
        P_1 = \bigg\{&|\uparrow\downarrow\dots\uparrow\downarrow\rangle\quad,\quad\big(\sigma_x^i\sigma_x^{i+1}\dots\sigma_x^{i+n}\big)|\uparrow\downarrow\dots\uparrow\downarrow\rangle\bigg\},\nonumber\\
        P_2 = \bigg\{&|\uparrow\uparrow\dots\uparrow\uparrow\rangle\quad,\quad\big(\sigma_x^i\sigma_x^{i+1}\dots\sigma_x^{i+n}\big)|\uparrow\uparrow\dots\uparrow\uparrow\rangle\bigg\},\nonumber\\
        P_3 = \bigg\{&\big|\downarrow\downarrow\dots\downarrow\downarrow\rangle,\nonumber
        \\&\quad\quad
        \big(\sigma_x^i\big)|\downarrow\downarrow\dots\downarrow\downarrow\rangle,\nonumber
        \\&\quad\quad\quad\quad
        \big(\sigma_x^i\sigma_x^j\big)|\downarrow\downarrow\dots\downarrow\downarrow\rangle\bigg\}.
    \end{align}
    
    Because the system is translation invariant, the zero-momentum sector is chosen as a numerical simplification. Under these constraints, each subspace $P$ has $N+1$ states each out of total Hilbert space dimension $\approx 2^N/N$.
    
    These basis states are each dressed by the variational AGP to create dressed boundary wall states: the hard boundary is softened by the dressing procedure to better describe the interacting quasiparticle excitations.

    \begin{figure*}
        \centering
        \includegraphics[width=0.49\linewidth]{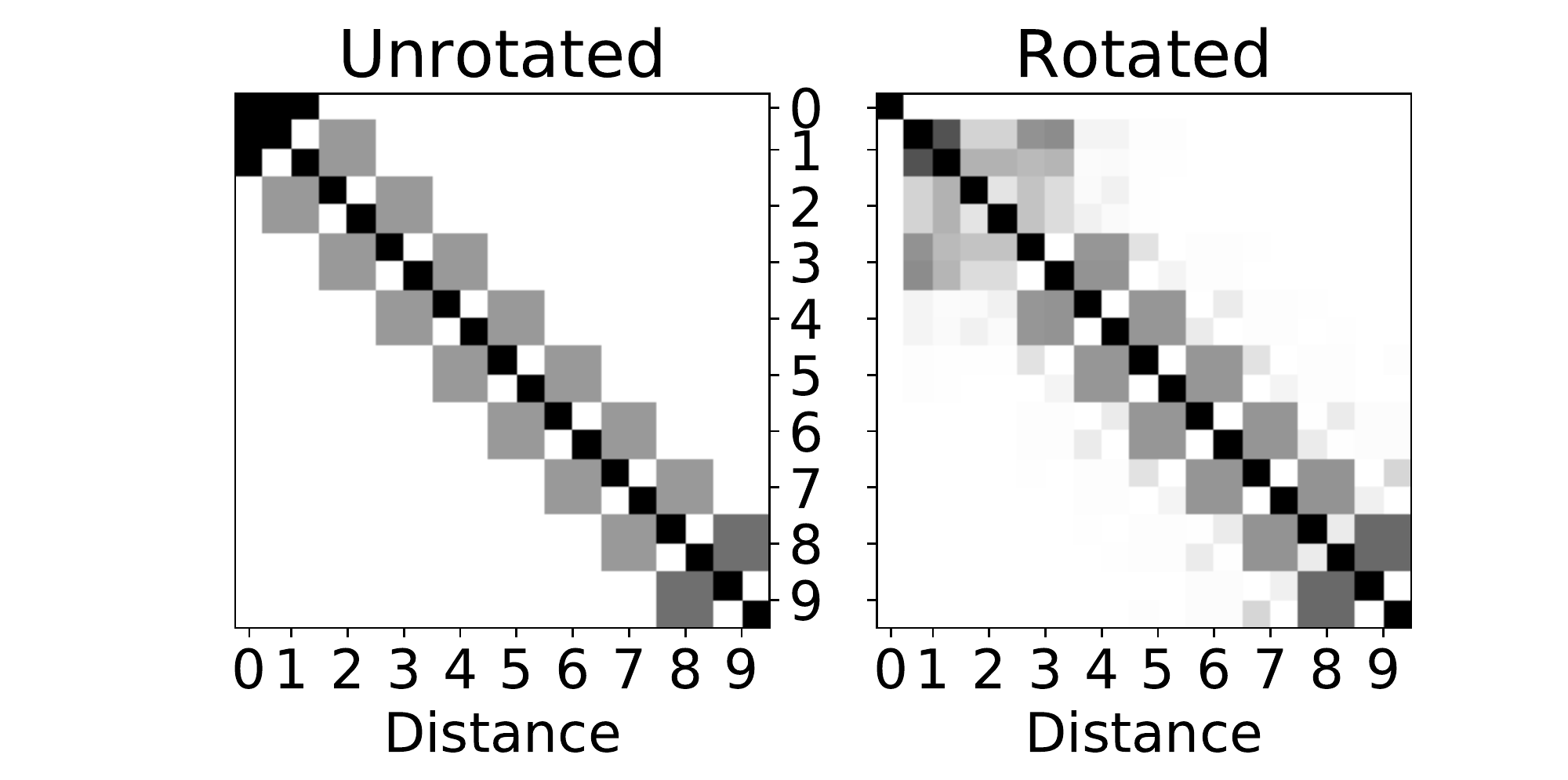}
        \includegraphics[width=0.49\linewidth]{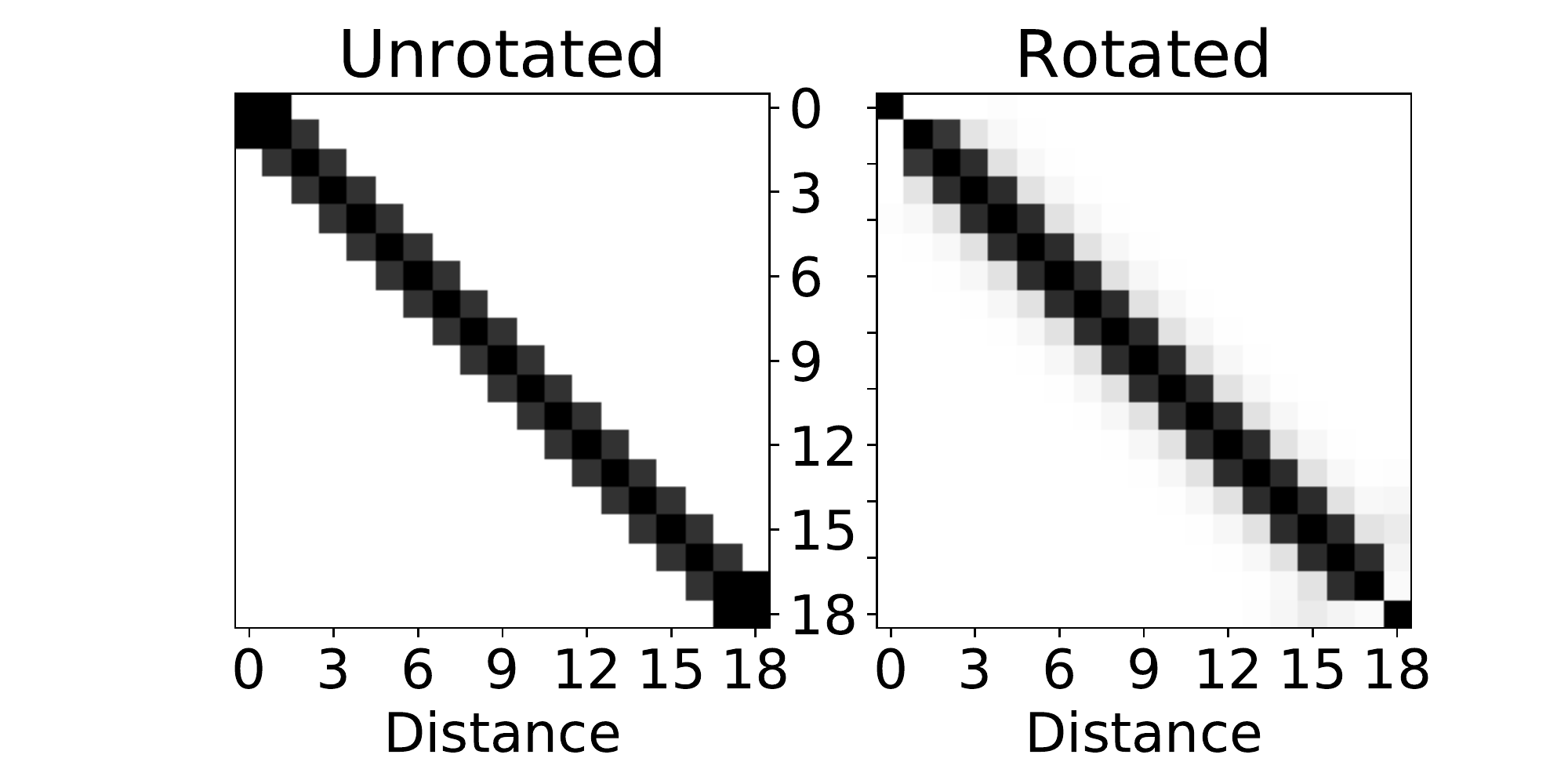}
        \includegraphics[width=0.49\linewidth]{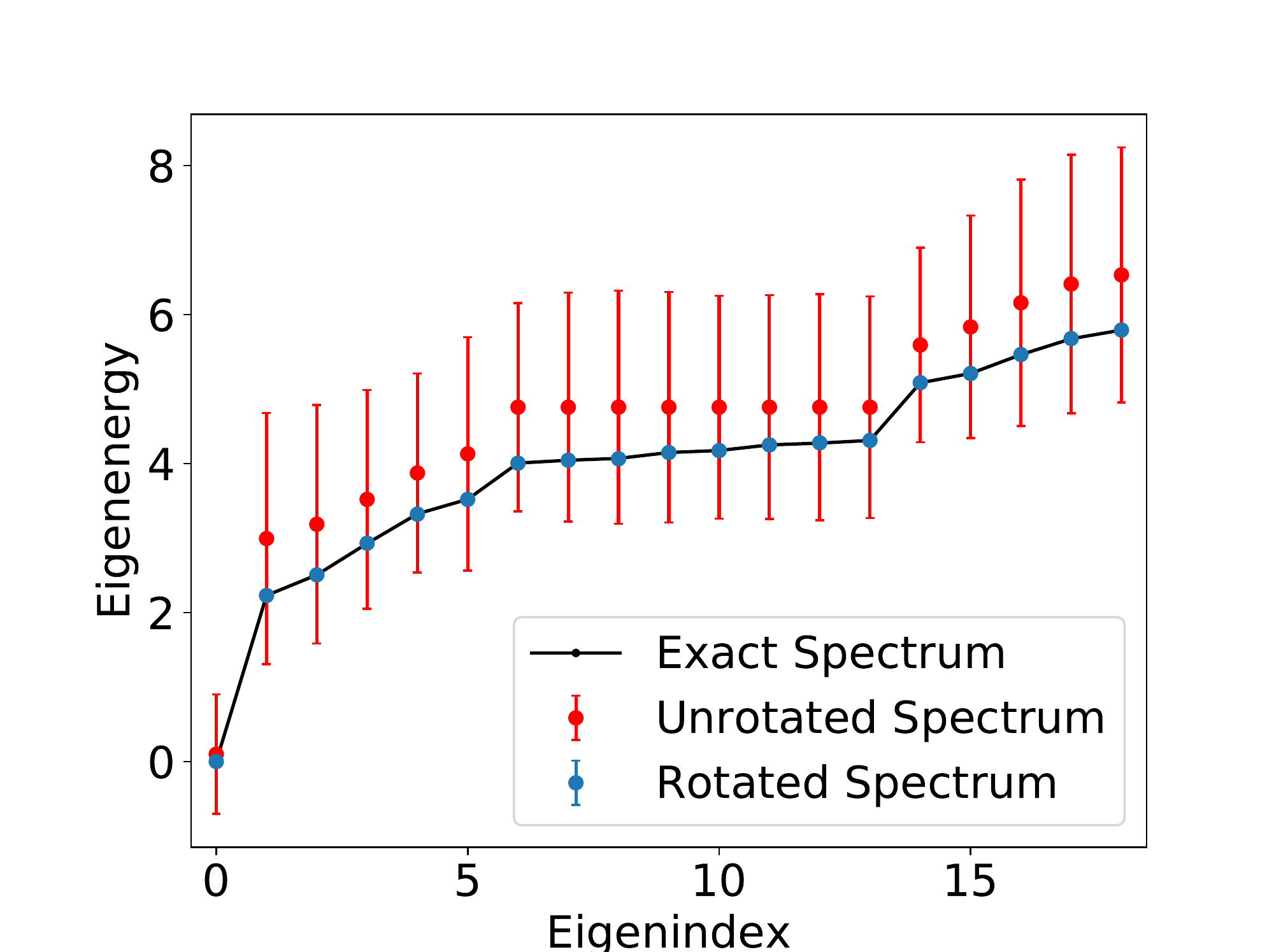}
        \includegraphics[width=0.49\linewidth]{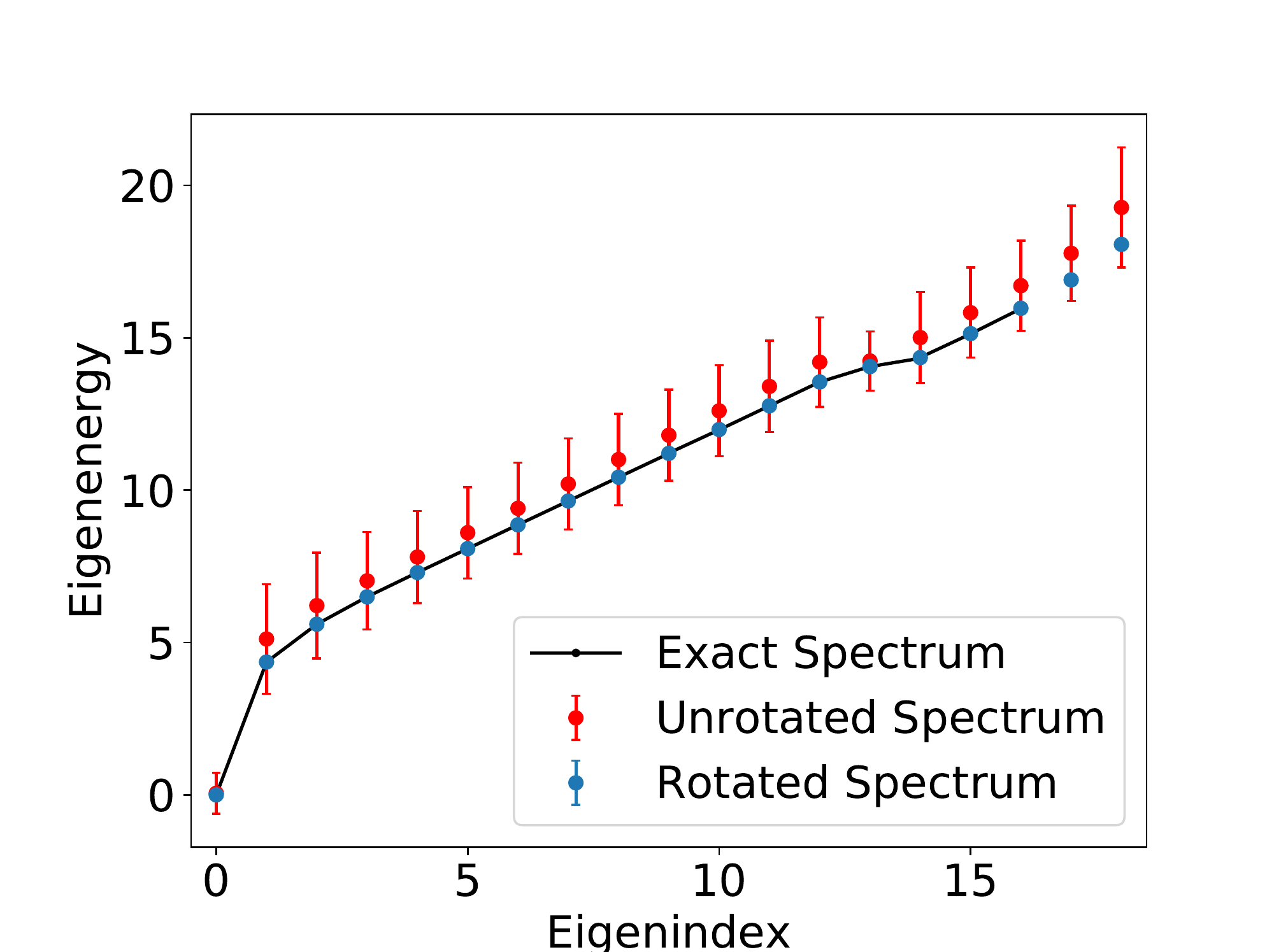}
        \caption{Eigenspectrum of the zero momentum mixed-field Ising model of Eq. \eqref{eq:base_hamiltonian}. Left is for 18-site AFM states ($J=+1$) while right is for 18-site FM states ($J=-1$). Top are matrix elements of the rotated and unrotated effective Hamiltonian indexed by boundary wall distance. Bottom is a comparison between the exact spectrum (computed numerically), the unrotated TSA spectrum (red), and the rotated spectrum (blue). Error bars are the energy variance of the approximate eigenstates.}
        \label{fig:TSA_spectrum}
    \end{figure*}
    
    \section{Approximate eigenstates and spectrum}
    
    As an explicit example, let us choose the parameters $h_x=0.4=h_z$, and coupling $J=\pm1$. These parameters are non-perturbative, in the sense $0.4$ is $\mathcal O(1)$ away from any simple point. For $J=-1$, the ground state is ferromagnetic, and the $h_z$ term acts as a constant attractive force between two boundary wall particles. This leads to ``meson" bound states of the two boundary walls \cite{Kormos2017}. For $J=+1$, the $h_z$ term does not change the AFM ground state energy  \cite{Ovchinnikov2003,Bonfim2019}; however it will affect energies of spin flips on up/down Ne\'el sublattices, which, from a band theory context, leads to two free particle species.
    
    \quad
    
    One can then go through the process of computing approximate eigenstates, as outlined in section \ref{sec:methods}, for these particular choices of subspaces. Here, Hamiltonians \eqref{eq:H1}, \eqref{eq:H2} are chosen starting from the FM and AFM subspaces, with an ansatz of all operators local to 3 sites plus Jordan Wigner strings, with 18 total sites in the 0 momentum sector. The form of the AGP is shown in Appendix \ref{app:AGPparameters}.
    
    Results for these parameters are shown in Fig. \ref{fig:TSA_spectrum}. Top plots the effective unrotated and rotated Hamiltonian, or equivalently the Hamiltonian in the projective and rotated projective subspaces, for AFM (left) and FM (right) excitations. It can be clearly seen that the rotated effective Hamiltonian becomes slightly more nonlocal: a dressed boundary wall of width 3 may hop to become width 5, for example. These effects are especially pronounced when the two boundary walls are close together, which is an indicator of a 2-particle interaction. When the two particles are far apart the Hamiltonian becomes independent of distance. 
    
    The spectrum is shown on the bottom plots of Fig. \ref{fig:TSA_spectrum}. Clearly, there is remarkable improvement over na\"ive TSA (red) with the unrotated basis, and the rotated version (blue) is almost identical to the exactly computed eigenspectrum (black). The error bars are the energy variance of the approximate eigenstates, as computed from Eq. \eqref{eq:energyvariance}.  Note that the exact eigenvectors are matched with approximate ones by choosing those which have maximum fidelity $|\langle E_n|E_m^\text{exact}\rangle|^2$; normally this value is $>0.9$.
    
    Importantly, the eigenstates are not necessarily all the lowest energy states. For example, two of the lowest-energy FM boundary walls (each with excitation energy $4.4J$) has a higher energy then a single FM boundary wall of width 6 (with excitation energy $8.7J$). This means it is energetically possible for the width-6 boundary wall to decay into two width-1 boundary walls, for example. However, the small energy variance of these dressed states indicates that such a process is almost completely suppressed.
    
    \quad
    
\begin{figure}
        \centering
        \includegraphics[width=\linewidth]{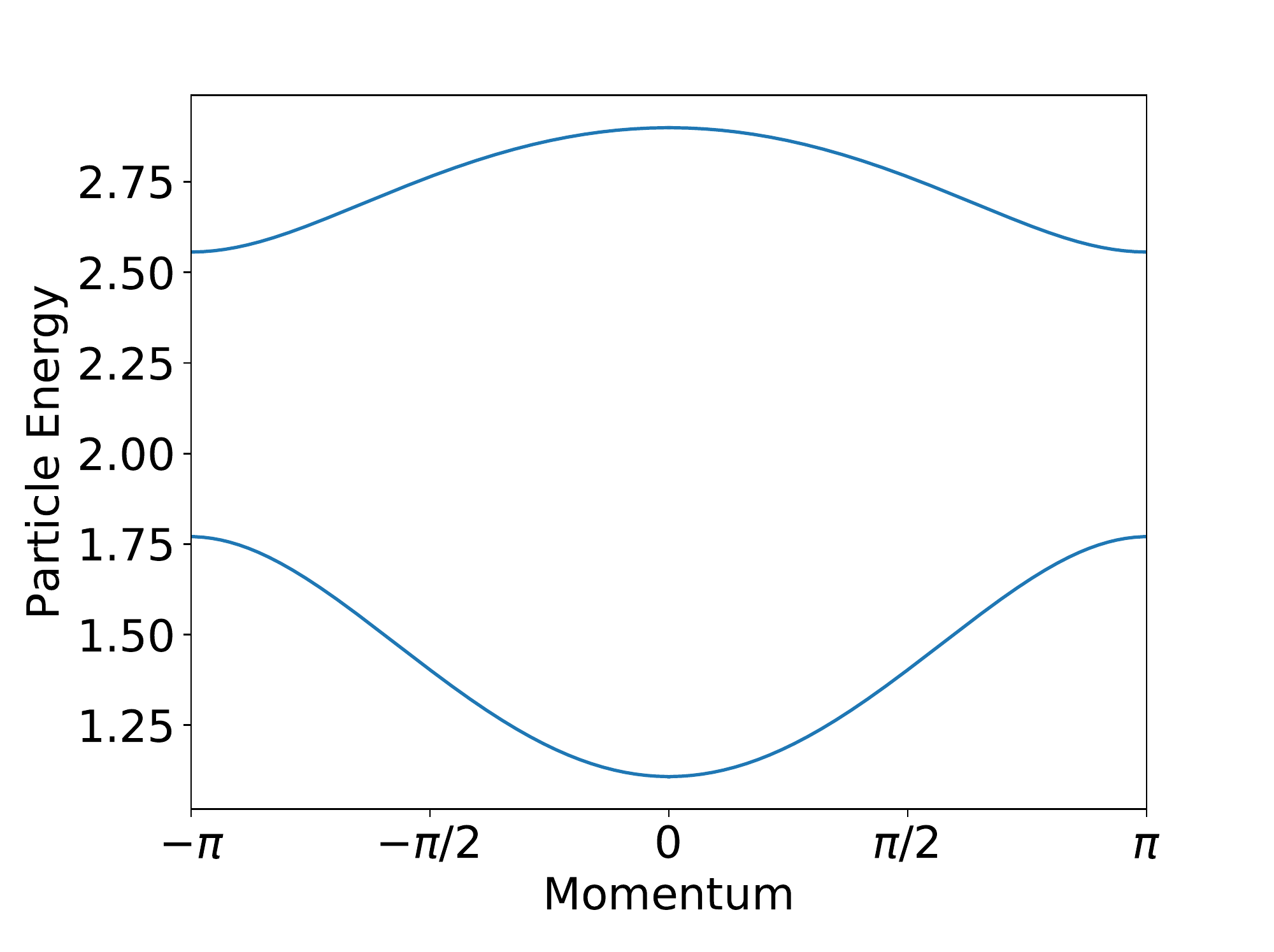}
        \caption{Dispersion relation of the two particle species on top of an AFM ground state. The 2-heavy particle energy lies above the 4-light particle continuum.}
        \label{fig:continuum_dispersion}
    \end{figure}

    Because the dressing is local, it is possible to take a continuum or large system size limit. Numerically, this is done by duplicating the dressed Hamiltonians of Fig. \ref{fig:TSA_spectrum} over thousands of sites: The $19\times 19$ matrix is extended to a $N\times N$ matrix, where the middle elements are the duplicated middle elements of the smaller matrix. Then, the eigensystem of that Hamiltonian is computed. Results for the continuum dispersion relation of excitations on top of the AFM ground state are shown in Fig. \ref{fig:continuum_dispersion}. There are two particle species which have mass $1.11$ and $2.56$. Like the meson case, the states are not necessarily the lowest-energy states: for example, 2 heavy particles can have equal energy to 4 light particles, and may potentially decay as such. Note that two light particles could not decay into 1 heavy particle, as that is disallowed by the particles being domain walls.

    \begin{figure}
        \centering
        \includegraphics[width=\linewidth]{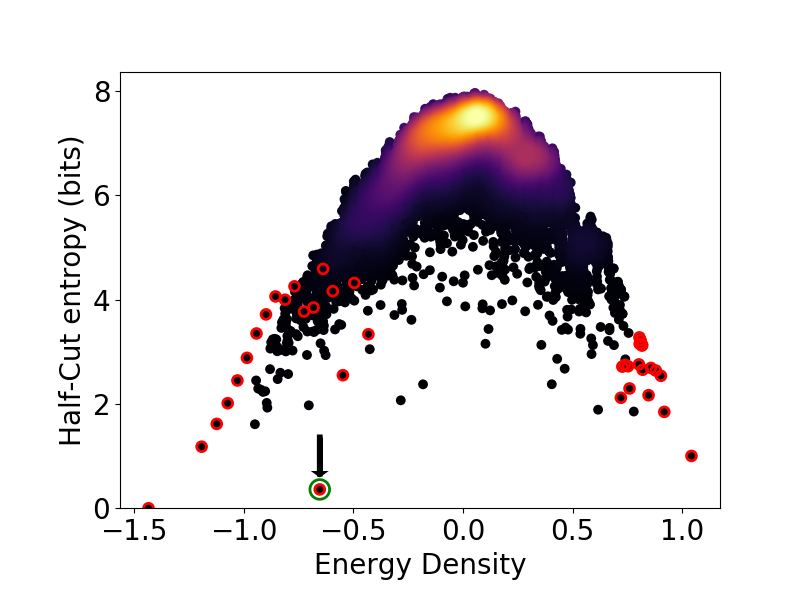}
        \includegraphics[width=\linewidth]{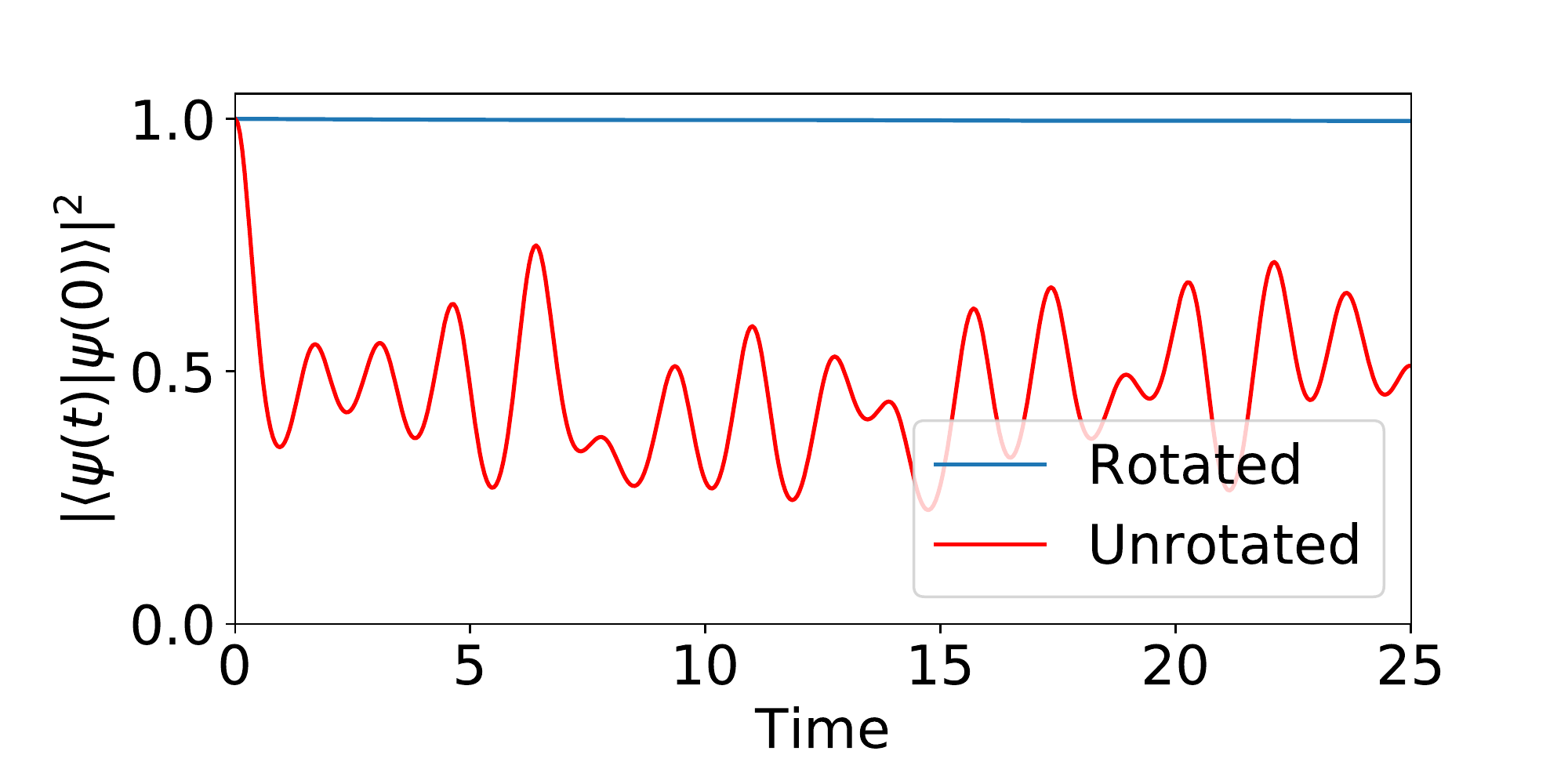}
        \caption{\textbf{Top:} Comparison of half-cut eigenstate entanglement entropy for the 18-site FM chain and 0 momentum. Red circles are the states with maximum overlap with the approximate eigenstates; left are dressed ferromagnetic states while right are dressed anti-ferromagnetic states. Note that high-energy states of the FM model are not the same as the low-energy states of the AFM model. Green circle and arrow indicates the dressed all-up state, which is a nonthermal state. \textbf{Bottom:} Fidelity of the dressed-all-up state with the initial state $|\langle E(t)|E(0)\rangle|^2$ which indicates the rotated all-up state is very close to an eigenstate, while the unrotated version is not, and is well preserved in time.}
        \label{fig:allup_state}
    \end{figure}
    
    These approximate eigenstates can be compared with the general bulk eigenstates, as is shown in Fig. \ref{fig:allup_state} Top. Here, all eigenstates in the 18-site FM model, 0-momentum sector (14,602 total) are computed exactly, and their half-cut entanglement entropy is found. Thermal states are extensively entangled states at finite energy density, while nonthermal states are weakly entangled and generally have zero or low energy density \cite{Turner2018a}. Red circles are the states which have maximal overlap with the approximate eigenstates: low energies are FM states while high energies are AFM states.
    
    One particular approximate eigenstate merits more study: the dressed all-up eigenstate, indicated by the green circle and arrow in Fig. \ref{fig:allup_state} Top. This is a ground state of $\sigma_z\sigma_z$, but the most excited state of $\sigma_z$; it has finite energy density given roughly by $2h_z$. But, in particular, it is an explicit example of a highly non-thermal state far from the edges of the spectrum \cite{Abanin2019,Turner2018a,Chengju2019,soonwon2019}. It is locally entangled with a half-chain entanglement entropy of $\approx 0.25$ bits. It has very high fidelity of $0.995$ with an exact eigenstate. Note that in the thermodynamic limit the dressed-all-up state is exponentially orthogonal to the original all-up state due to the finite local rotation.

    \subsection{Quasiparticle Lifetimes}
    
    The energy variance of these approximate eigenstates takes special meaning when they can be interpreted as dressed particles. In this case, the energy variance gives a lower bound on the quasiparticle lifetime. For an approximate particle eigenstate $|E_n\rangle$, the time-dependent state overlap under second order perturbation theory is
    
    \begin{align}\label{eq:state_fidelity}
        \big|\langle E_n|E_n(t)\rangle\big|^2
        \approx&1 - \frac{t^2}{\tau^2} + \mathcal O(t^4),
    \end{align}
    where $\tau^{-2} = \Delta^2 = \langle H^2\rangle - \langle H\rangle^2$ is the energy variance of the state. In other words, the characteristic time for an (eigen)state of some particles to decay into some other particle state is given by the energy variance. This timescale is very crude as it assumes all other states have the same energy: a more refined timescale can be computed using the Fermi golden rule \cite{Sachdev2011a} for the dressed states, but is not generally possible without a priori knowledge of the energy of the other states. As such, the energy variance serves as a lowest bound on (inverse) quasiparticle lifetime. 
    
    As an explicit example of these timescales, a dressed single flipped down spin on a FM ground state, which corresponds to the lowest energy meson excitation, has a characteristic lifetime of $\tau=110$, far longer than any local timescale. Excitations on top of an AFM ground state have lifetimes in excess of $\tau>60$. The dressed all-up state has a lifetime of $\tau=53$. These lifetimes are longer when adding more parameters to the variational AGP. Explicit time dynamics of Eq. \eqref{eq:state_fidelity} for this dressed all-up state is shown in Fig. \ref{fig:allup_state} Bottom. Clearly, it is much closer to an eigenstate than expected, as it is close to 1 at all times. This further indicates the genuineness of this nonthermal eigenstate, especially when compared to the undressed version of the same. Note that the undressed up state is exponentially orthogonal in system size from the dressed up state, due to the finite rotation.

    One application of such a dressed all-up state is for information protection in quantum systems. For a classical Hamiltonian, the all-down [ground] state may be labeled as a logical 0, while the all-up state is labeled as a logical 1. An X-field will generally change these two states, destroying the encoded bit. If this bit is instead encoded in the dressed nonthermal states (the all down ground state, and the all up nonthermal state), they are much more stable, encoding the information for a much longer time by suppressing transitions.
    
    \quad
    
    \subsection{Quasiparticle parameter dependence}\label{sec:QP_parameter_dependence}

    \begin{figure}
        \centering
        \includegraphics[width=\linewidth]{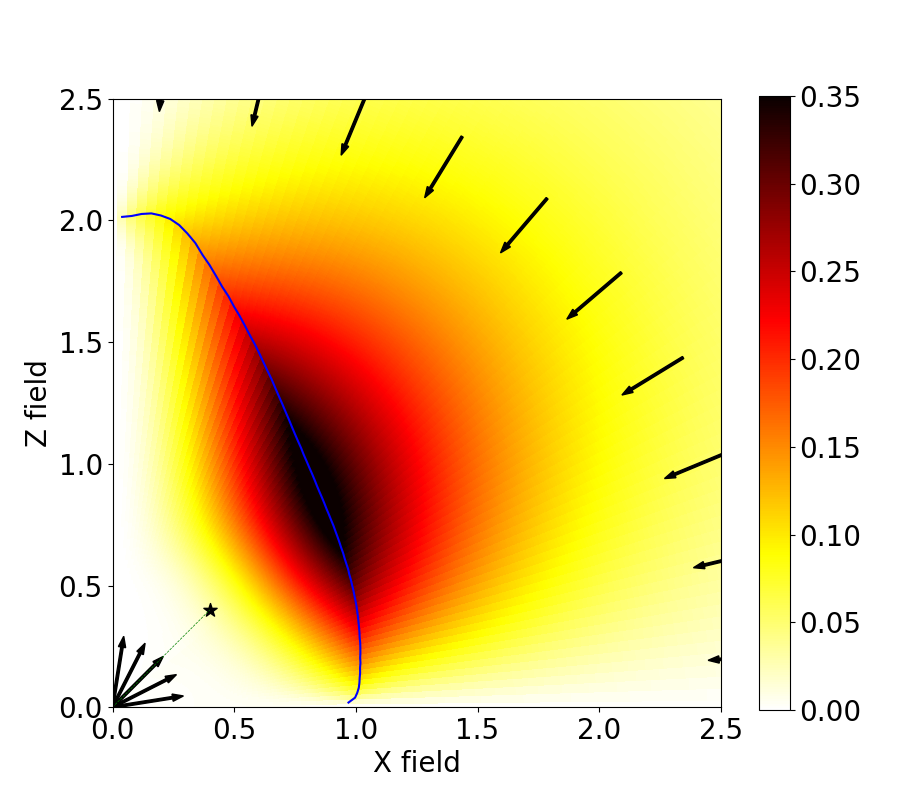}
        \includegraphics[width=\linewidth]{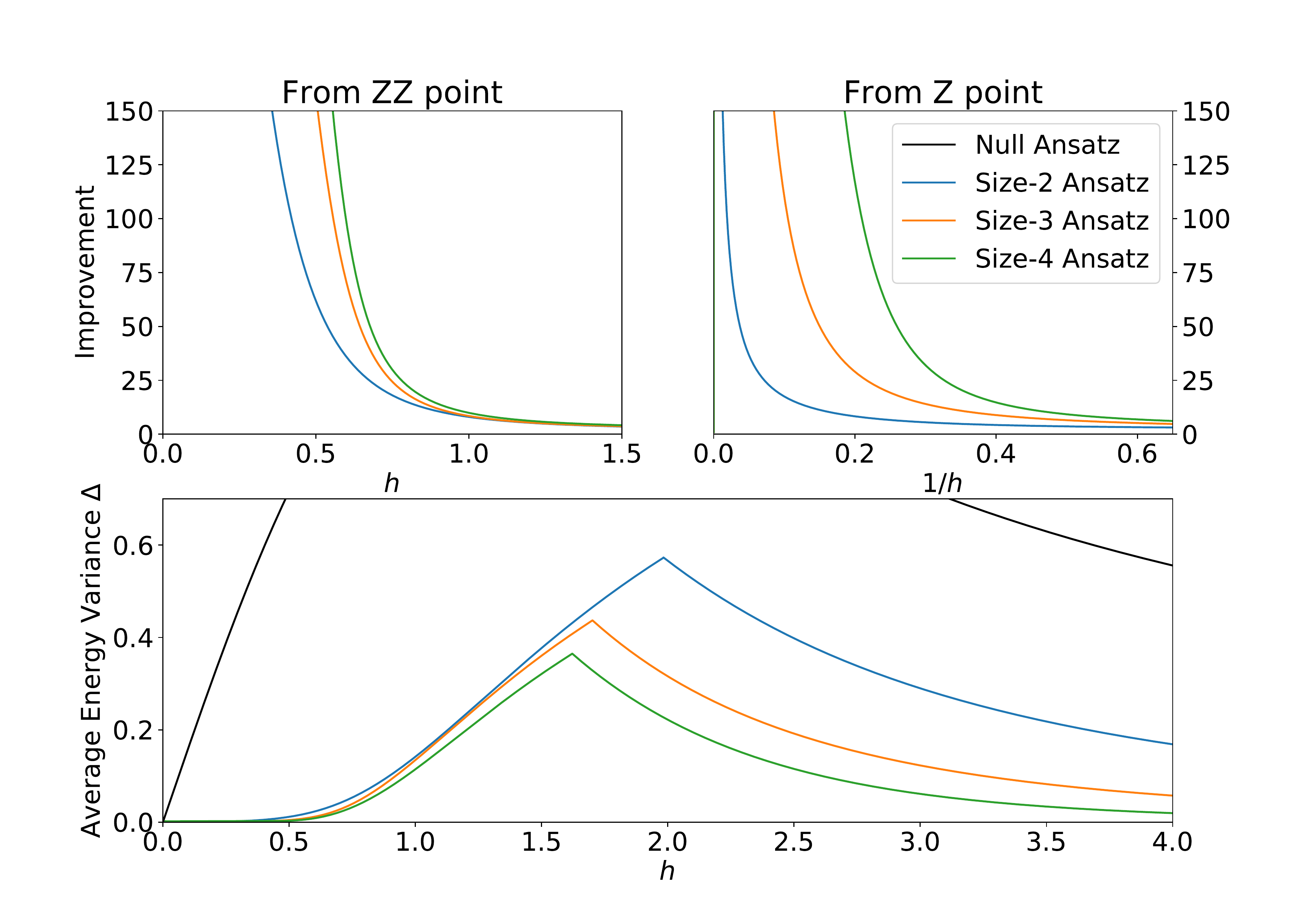}
        \caption{Results for a 3-site VGP ansatz for the 14-site TLI Model. \textbf{Top:} Average inverse 2-particle lifetime $\Gamma$ or equivalently average energy variance. Blue line indicates transition between different directions. Star is the $(0.4,0.4)$ point studied more in-depth. Arrows indicate direction of dressing.  White and yellow indicate areas with a good quasiparticle description. \textbf{Bottom:} Average energy variance in the direction $\pi/4$ from vertical. Middle is improvement from the undressed subspace. Below $h\approx0.6$, the error is vanishingly small. \textbf{Middle:} Energy variance improvement compared to the Null ansatz $\Gamma/\Gamma_0$ along the $h_x=h_z$ line. As the ansatz size increases, so too does the improvement, as expected.}
        \label{fig:energy_fluctuation}
    \end{figure}

    A general measure of the quasiparticle lifetime within a particular subspace is given by their normalized average inverse lifetime, or equivalently average energy variance
    
    \begin{equation}
        \Gamma = \frac{1}{\sqrt{\mathcal D_p}\sqrt{1 + h_x^2 + h_z^2}}\sqrt{\sum_n\Delta_n^2}.
    \end{equation}
    
    $\Gamma$ equivalently is the block-off-diagonal weight of the rotated subspace $\tilde P$ (See Appendix \ref{app:Eflucs}). A small value indicates a good block-diagonalization procedure with well-defined quasiparticles within the subspace. A large value indicates a failure to block diagonalize the Hamiltonian.
    
    This error can be computed for various values of $h_x$ and $h_z$ by evolving with parameterized Hamiltonians \eqref{eq:H1} and \eqref{eq:H3}, computing approximate eigenvalues, then computing their normalized average energy variance $\Gamma(h_x,h_z)$. Results are shown in Fig. \ref{fig:energy_fluctuation}, for a 3-site ansatz and 14 sites. States are dressed from one of two directions. One is $H_1(\mu)$, dressing 2-particle AFM boundary wall states out from the $\sigma_z\sigma_z$ only point, indicated by the radial arrows in the bottom left. The other is $H_3(\mu)$, dressing 1 and 2-particle spin-flip states from the $h_x\sigma_x + h_z\sigma_z$ only point(s), indicated by the arrows pointing radially inwards.
    
    In the region where $h_x$, $h_z$ is small, the error from dressing boundary walls is enormously low. With no dressing, the error grows linearly in $|h|$, while with dressing, the error grows sub-linearly, which indicates that the dressing is exact asymptotically. In fact, this dressing accumulates very small errors even for non-perturbative values of $|h|$, as shown in Fig. \ref{fig:energy_fluctuation} Bottom, which is dressing along the $h_x=h_z$ line. This indicates that in the white areas, there is a good effective quasiparticle description of the low energies of this otherwise quantum chaotic model, described by dressed boundary wall particles.
    
    Although it is not generally so, the error accumulates monotonically with increasing $|h|$. This means that at some critical value, the dressing going \textit{outwards} from the $\sigma_z\sigma_z$ point will have a larger error then the dressing going \textit{inwards} from the $(h_x,h_z)\to\infty$ point. At this boundary, the best description of quasiparticles changes from dressed pairs of \textit{boundary walls}, to dressed \textit{spin flips}. This does not mean that there is no effective description of certain states in terms of quasiparticles: there could be some other subspace (say, of doubly-flipped spins) and other path through parameter space which gives a better quasiparticle picture in that the energy variance is smaller.

    This cross-over point may be an indicator of an interacting phase transition. Around $h_z=0,h_x=1$, which is the transverse Ising phase transition, it has been found that local variational adiabatic dressing begins to fail \cite{Polkovnikov2017}. This finding is now extended to the interacting case: the cross-over gives a rough region where the interacting critical point may occur, as a local AGP fails to reproduce the long-range entanglement of a critical ground state \cite{hastings2007}. With increasing ansatz size, this point decreases in total error, and shifts in critical parameter (see Fig. \ref{fig:energy_fluctuation} Bottom) which may eventually converge to some particular value, indicating the interacting critical point. This idea is backed up by the convergence in the non-interacting limit: For $h_z$ small, the crossover is around the $h_x\approx 1$ transverse Ising critical point. Similarly, for $h_x$ small, the crossover is around $h_z\approx 2$, which is the first-order phase transition in $h_z$ to change the ground state from AFM to polarized \cite{Simon2011}.

    \section{Local almost-conserved operators}

    \begin{figure*}
        \centering
        \includegraphics[width=0.95\linewidth]{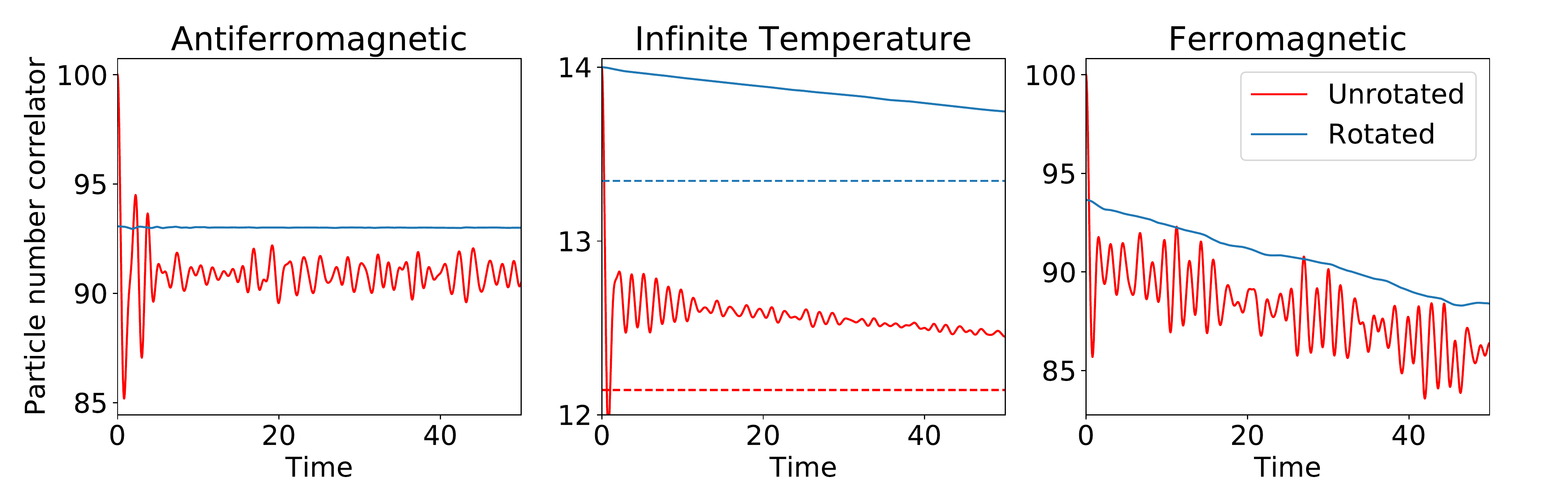}
        \caption{Symmetric time correlation function of dressed (blue) and undressed (red) particle number of Eq. \eqref{eq:Ntilde} and Eq. \eqref{eq:Nnotilde} for a 2 particle AFM subspace (left), 2 particle FM subspace (right) and all states (middle), and 14 sites. Dashed are the infinite temperature long-time values.}\label{fig:particlenumber}
    \end{figure*}

    Given some set of approximate local eigenvectors $\{|E_n\rangle\}$ generated by this adiabatic dressing scheme, it is a relatively simple procedure to construct approximately conserved local operators. An operator is conserved if it commutes with the Hamiltonian, or equivalently if it is constructed from eigenstates of the Hamiltonian. Given approximate eigenstates, one should then be able to compute approximately conserved quantities. Conserved operators are defined as
    
    \begin{equation}\label{eq:conserved_operator_defn}
        \mathcal O = \sum_n O_n |E_n\rangle\langle E_n|
    \end{equation}
    where $O_n$ are the eigenstates of the operator, and$\{|E_n\rangle\}$ are exact eigenstates. An operator of this form has the property that the symmetric time correlation function is conserved (for all initial states)
    
    \begin{equation}\label{eq:cons_symcorr}
        \frac{\langle \{\mathcal O(t),\mathcal O(0)\}\rangle}{2} = \text{Constant}.
    \end{equation}

    There are two ways to construct approximate versions of these operators. The first way is to explicitly use Eq. \eqref{eq:conserved_operator_defn} using only the subspace of dressed eigenstates which were directly computed. In this case, the sum is of size $\mathcal D_P$, the subspace size, as opposed to $\mathcal D$, the total Hilbert space size. Due to the global projective structure of particle excitations on top of a ground state, the resulting operator is not necessary local. However, this may be implemented with some local operator plus post-selection of states.
    
    In the case of such an operator directly constructed from approximate eigenstates, the symmetric correlation function is not conserved in time. Under perturbation theory, the characteristic timescale is given by a weighted-average energy variance (see Appendix \ref{app:symmcorr} for derivation)

    \begin{align}\label{eq:ncons_symcorr}
        \frac{\langle\{O(t),O(0)\}\rangle}{2} &= \langle O^2\rangle\bigg(1 - \frac{t^2}{\tau_O^2}\bigg),\nonumber\\
        \frac{1}{\tau_O^2} &\equiv \frac{\sum_n \rho_nO_n^2\Delta_n^2}{\sum_n \rho_nO_n^2}.
    \end{align}
    
    Here, $\rho_n=\langle E_n|\rho|E_n\rangle$ is the density matrix for the expectation value, assumed to be diagonal, and $\Delta_n$ is the energy variance of the $n$th approximate eigenstate. For good eigenstates with low energy variance, the decay timescale can be very long.
    
    \quad
    
    The second way to construct approximately conserved operators is to dress conservation laws of the simple system $H(0)$ with the unitary. Conservation laws, such as particle number and particle current, are constructed from simple eigenstates in the form of Eq. \eqref{eq:conserved_operator_defn}, with particular choice of $O_n$, and are generally local \cite{Calabrese2012a,Karbach1998}. Then, the dressed operator is constructed from states better resembling eigenstates
    
    \begin{align}
        \mathcal O &= U\bigg(\sum_n O_{n} \big|E_n(0)\big\rangle\big\langle E_n(0)\big|\bigg)U^\dagger\\
        &\Updownarrow\nonumber\\
        \mathcal O &= U\mathcal O_0U^\dagger.
    \end{align}
    
    Importantly, the eigenstates of operator $\mathcal O$ are not necessarily the same as the constructed approximate eigenstates, as it is missing the re-diagonalization step of Eq. \eqref{eq:re-diagonalization2}. The resulting operator is quasi-local, and approximately conserved \cite{serbyn2013,serbyn2014}. As the ansatz span is increased, the AGP approaches the exact one, resulting in better approximate eigenstates and a better-conserved operator, at the expense of it becoming more and more nonlocal.
    
    One such conserved operator for the mixed-field Ising model is the dressed total particle number $N = U\big(\sum_i\sigma_z^i\sigma_z^{i+1}\big)U^\dagger$ which (up to a constant) counts the number of boundary walls in the system. For $H_1$ and $H_2$, this is also the initial Hamiltonian; thus one would expect that the dressed operator should also approximate dressed versions of particle number eigenstates.
    
    For $h_x=h_z=0.4$ and $J=+1$, the dressed particle number operator becomes

    \begin{align}
        N_0 = \sum_i\sigma_z^i\sigma_z^{i+1}\quad\quad\quad&\label{eq:Nnotilde}\\
        \Downarrow\quad\quad\quad\quad\quad&\nonumber\\
        N = \sum_i0.9530\hat\sigma^{i}_z\hat\sigma^{i+1}_z+\label{eq:Ntilde}\\\nonumber
        0.2135\hat\sigma^{i}_x+0.1927\hat\sigma^{i}_z\hat\sigma^{i+1}_x\hat\sigma^{i+2}_z+\\\nonumber
        0.0616\hat\sigma^{i}_z\hat\sigma^{i+1}_x\hat\sigma^{i+2}_x\hat\sigma^{i+3}_z+\\\nonumber
        -0.0398(\hat\sigma^{i}_z\hat\sigma^{i+1}_x+\hat\sigma^{i}_x\hat\sigma^{i+1}_z)+\\\nonumber
        -0.0243\hat\sigma^{i}_y\hat\sigma^{i+1}_y+\\\nonumber
        0.0211\hat\sigma^{i}_z\hat\sigma^{i+1}_x\hat\sigma^{i+2}_x\hat\sigma^{i+3}_x\hat\sigma^{i+4}_z+\\\nonumber
        -0.0164(\hat\sigma^{i}_x\hat\sigma^{i+1}_x\hat\sigma^{i+2}_z+\hat\sigma^{i}_z\hat\sigma^{i+1}_x\hat\sigma^{i+2}_x)+\\\nonumber
        0.0098\hat\sigma^{i}_z+\\\nonumber
    +\dots\nonumber
    \end{align}
    where ellipsis represent the more and more nonlocal terms of the operator. As can be seen, this operator is approximately local, with dominant terms coming from 1, 2, and 3-spin terms. One can then compute the symmetric correlation function in the initial undressed subspace of two particles to see its conservation
    
    \begin{equation}
        \tr{\mathcal P_0\{N(t),N(0)\}} = C(t).
    \end{equation}
    
    Results are shown in Fig. \ref{fig:particlenumber} for AFM and FM states, as well as infinite temperature typical states \cite{Goldstein2006}. For comparison, the undressed operator is also shown in red. The conserved operator for AFM states is almost stationary in time, while the undressed version is not. The infinite temperature timescale can be computed analytically as 
    
    \begin{align}
        \frac{\tr{\big([N,H]\big)^2}}{2\tr{N^2}}=&\tau_N^{-2}=50.93^{-2},\\
        \frac{\tr{\big([N_0,H]\big)^2}}{2\tr{N_0^2}}=&(\tau_{N_0})^{-2}=1.25^{-2}.
    \end{align}
    
    Even for an infinite temperature state, this quasilocal dressed operator gets a factor of 40 improvement in the characteristic decay timescale. This indicates that dressed quasiparticle excitations may persist in this interacting model \emph{even at infinite temperature}, and that this model is ``closer to integrable" then one might expect.

    \section{Conclusion}

    The existence of good approximate dressings have some curious implications. Even if a model system is not necessarily integrable or exactly solvable, that does not mean that there are no local long-lived symmetries and conservation laws. Indeed, if such a model is close by to an integrable point, a conservation law of the integrable model can be ``dressed" by a unitary generated by the approximate local adiabatic gauge potential to restore the symmetry approximately in a now quasi-local operator. Approximate eigenstates may be computed in a similar manner: simple particle excitations of the integrable point can be dressed by the approximate AGP to construct long-lived quasiparticle excitations of the interacting point. These new dressed states need not be low energy states and in fact may be used to construct finite energy density low-entanglement nonthermal states, as demonstrated in the dressed-all-up state of the mixed-field Ising model.

    Similar studies have been done to compute low energy phenomenology of the Meson case, most predominantly in recent work by \cite{Konik2019,Robinson2019} using a truncated spectrum approach (TSA). These numerical diagonalization procedures are functionally equivalent except that here the projective subspace is first rotated by the variational AGP, leading to a subspace closer to the exact eigenstates. While this work uses discrete lattices, generalizations to continuous theories is an interesting future direction.

    The restoration of approximate symmetries and construction of quasiparticle excitations in interacting models puts a new perspective on integrability breaking. Instead of reevaluating a Hamiltonian for every new point in parameter space, one can instead compute properties and approximate symmetries based on nearby Hamiltonians with a potentially simpler structure. This ``closeness" is defined in the sense of being able to compute a good approximate AGP along some path between the simple Hamiltonian and interacting one, not in the sense of perturbative parameter changes. Certain perturbations away from integrability may rapidly destroy any local conservation laws, if there exists no good local approximate AGP. Other perturbations, while still breaking integrability, may still admit quasi-local conservation laws, nonthermal states, and quasiparticles, if there does exist a good local approximate AGP.
    
    These unitary rotations restoring approximate integrability are similar in spirit to canonical transformations in KAM theory~\cite{Brandino2015}: integrability may be approximately restored for particular subsets of initial conditions of particular integrability-broken systems via the unitary rotation (eg canonical transformation) of conserved quantities. Whether this approach to stability of quantum integrable systems can be made more concrete remains to be seen but these variational local dressings may be a step towards a general theory in that direction.

\acknowledgements

We would like to thank Artem Rakcheev, Sho Sugiura and Pieter Claeys for useful discussions. J.W. and A.P. were supported by
NSF Grants No. DMR-1813499 and No. AFOSR FA9550-16-
1-0334.

    \newpage

    \appendix

    \section{Hastings 2005 Eq. 17 to Eq. \eqref{eq:regularized_AGP}}\label{app:hastings05}
    
    This section serves as a derivation of Eq. \eqref{eq:regularized_AGP} from \cite{Hastings2005} Eq. 17. Hastings defines the rotation $\tilde V(s)$ (analogously $U^\dagger$) as
    
    \begin{equation}
        \tilde V(s) = \mathcal S'\exp\bigg\{-\int_0^sds'\int_0^\infty d\tau e^{-(\tau/\tau_q)^2/2}[\tilde u^+_{s'}(i\tau)-\text{H.c.}]\bigg\}.
    \end{equation}
    
    Here, $\mathcal S'$ is parameter ordering (analogously $\mathcal T$) for parameter $s'$ (analogously $\mu$). The object $\tilde u^+_{s'}(i\tau)$ is defined as
    
    \begin{equation}
        \tilde u^{\pm}(\pm i\tau) \equiv \frac{1}{2\pi}\int_{-\infty}^\infty dt \frac{[\partial_s H_s](t)e^{-(t/\tau_q)^2/2}}{\pm it + \tau}
    \end{equation}
    with $[*](t)$ denoting time evolution with respect to instantaneous Hamiltonian $H_{s'}$ (analogously $H(\mu)$). Substituting this into the inner integrand and simplifying by integrating over $\tau$ yields
    
    \begin{widetext}
    \begin{multline}
        \int_0^\infty d\tau e^{-(\tau/\tau_q)^2/2}[\tilde u^+_{s'}(i\tau)-\text{H.c.}]\quad=\quad
        \frac{1}{2\pi}\int_{-\infty}^\infty dt [\partial_s H_s](t)e^{-(t/\tau_q)^2/2}\int_0^\infty d\tau e^{-(\tau/\tau_q)^2/2}\bigg[\frac{2it}{t^2+\tau^2}\bigg]\\\quad\\
        =\quad\frac{1}{2\pi}\int_{-\infty}^\infty dt [\partial_s H_s](t)e^{-(t/\tau_q)^2/2}\bigg[-i\pi e^{(t/\tau_q)^2}\text{erfc}\Big(\Big|\frac{t}{\sqrt{2}\tau_q}\Big|\Big)\text{SGN}(t)\bigg]\quad=\quad
        \frac{-i}{2}\int_{-\infty}^{\infty}dt [\partial_s H_s](t)\text{erfc}\Big(\Big|\frac{t}{\sqrt{2}\tau_q}\Big|\Big)\text{SGN}(t).\nonumber
    \end{multline}
    \end{widetext}
    Up to a trivial factor of $\sqrt{2}$ which can be absorbed by the regularization time $\tau_q$, this is the expression in Eq. \eqref{eq:regularized_AGP}.

    \section{Off-diagonal matrix elements and energy variance}\label{app:Eflucs}
    
    The average energy variance of approximate eigenstates analogously gives the average block-off-diagonal elements in the Hamiltonian, as claimed in section \ref{sec:QP_parameter_dependence}. This Appendix serves to elaborate on this point.
    
    When computing an effective Hamiltonian within a rotated subspace, the procedure is an analogous one to a Schrieffer-Wolff transformation: a unitary rotation block diagonalizes some Hamiltonian into a subspace $P$ and complement $Q$. A measure of the quality of this diagonalization is the average strength of the off-diagonal elements: zero strength means exact block diagonalization, while nonzero strength means approximate diagonalization. The average energy variance is defined as
    
    \begin{align}
        \Gamma^2 &= \frac{1}{\mathcal D_P}\sum_n^p\langle E_n|H^2|E_n\rangle - \Big(\langle E_n|H|E_n\rangle\Big)^2,\\
        \Gamma^2&=\frac{1}{\mathcal D_P}\sum_n^p\langle E_n| H\Big(|q\rangle\langle q| + |E_{p'}\rangle\langle E_{p'}|\Big) H|E_p\rangle\nonumber\\
        &\quad\quad\quad\quad- \Big(\langle E_n|H|E_n\rangle\Big)^2,\\
        \Gamma^2&=\frac{1}{\mathcal D_P}\sum_{nq}\Big|\langle E_n|H|q\rangle\Big|^2.
    \end{align}
    
    Step 2 inserts the identity, for complete set of states $|q\rangle\in Q$, and complete set of states $|E_p\rangle \in P$, while step 3 simplifies using the fact that $|E_n\rangle$ are eigenstates of the effective Hamiltonian within subspace $P$. Because $|E_n\rangle$ is a complete set of states in $P$ and similarly for $Q$, the sum is then over all off-block-diagonal matrix elements, giving an average off diagonal strength.

    \section{Path Dependence}\label{app:pathdependence}
      
    \begin{figure}
        \centering
        \includegraphics[width=\linewidth]{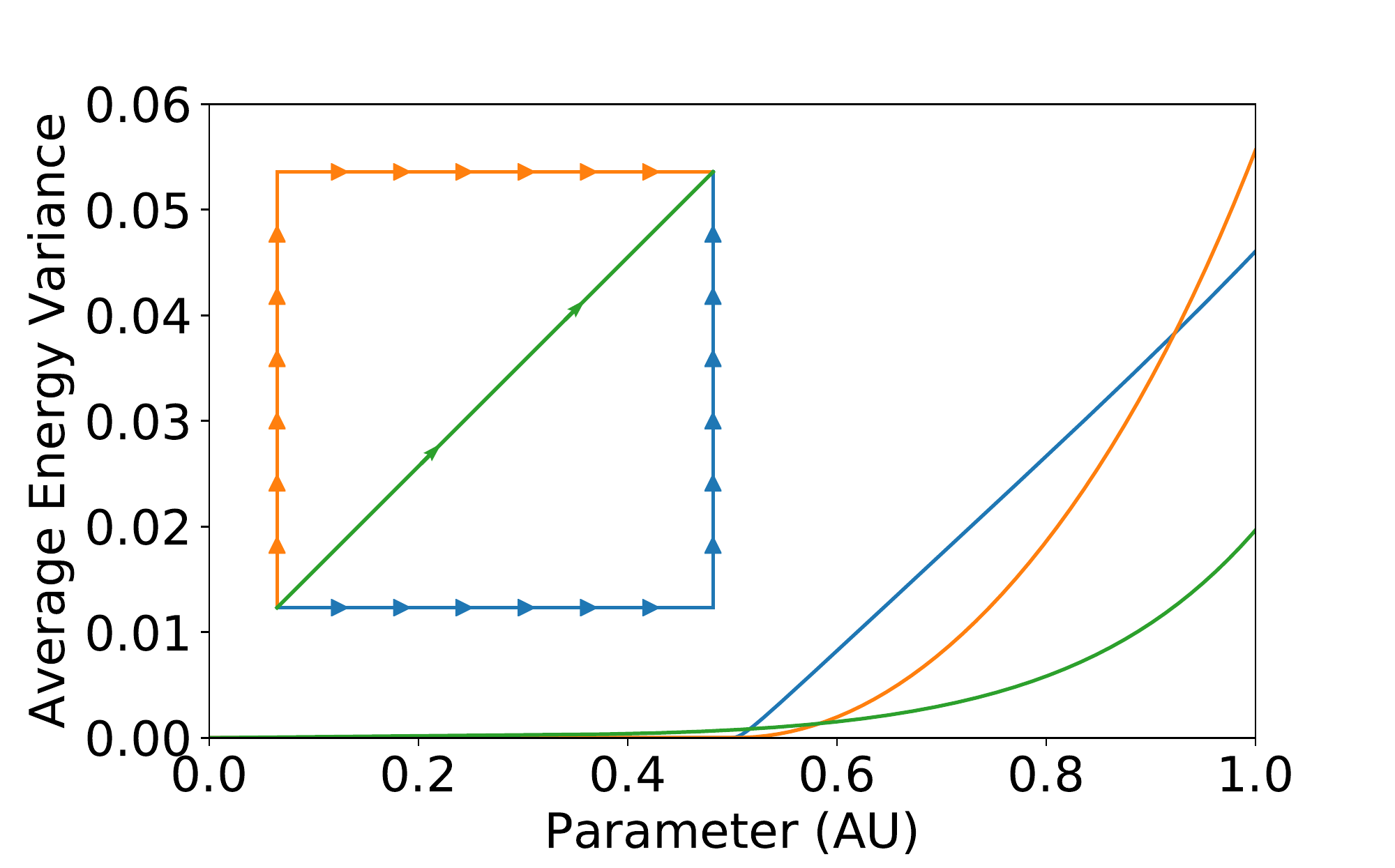}
        \caption{Three paths to get to the (0.4,0.4) point: Along X then along Z (blue); along Z then along X (orange) and diagonally (green). Energy variance is along each point on the path; the paths turn at parameter value $0.5$.}
        \label{fig:path_dependence}
    \end{figure}

    While this work chooses one particular path to the (0.4,0.4) point, there exist many other options, which may give better or worse performance. Here two additional paths are compared to the diagonal path. The second goes up in $h_z$, for which the gauge potential is zero, and then right in $h_x$. The third goes right in $h_x$ as the transverse Ising model for which the gauge potential is exact, and then up in $h_z$ (See inset in Fig. \ref{fig:path_dependence}). This directionality is shown in Fig. \ref{fig:path_dependence} for 14 sites, and a size-3 ansatz. The diagonal path ends with an average energy variance of $\approx 0.02$, while the other paths have variance of $\approx 0.04$, a factor of two improvement.

    \section{Symmetric correlators and almost conserved quantities}\label{app:symmcorr}
    
    This appendix serves as a derivation of Eq. \eqref{eq:ncons_symcorr}. Suppose some set of $\mathcal D_P$ approximate eigenstates $|E_n\rangle$ diagonalized within some subspace $P$, and operator $\mathcal O = \sum_n O_n|E_n\rangle\langle E_n|$ for Hamiltonian $H$. For simplicity, let us choose a subspace $[\rho,O]=0$ or equivalently $\rho = \sum_n \rho_n|E_n\rangle\langle E_n|$. In this case, the symmetric correlation function may be equivalently written as
    
    \begin{equation}
        \frac{\tr{\rho \{\mathcal O(t),\mathcal O(0)\}}}{2} = \tr{\rho \mathcal O(t) \mathcal O}.
    \end{equation}
    
    Next, expand the operator to second order in a BCH series
    
    \begin{equation}
        \mathcal O(t)\approx \mathcal O + it[H,\mathcal O] - \frac{t^2}{2}\big[H,[H,\mathcal O]\big] + \dots
    \end{equation}
    and substitute back in. The first order term is zero via trace identities. What remains is
    
    \begin{equation}
    \frac{\tr{\rho \{\mathcal O(t),\mathcal O(0)\}}}{2}\approx \langle \mathcal O^2\rangle - \frac{t^2}{2}\tr{\rho\big[H,[H,\mathcal O]\big]\mathcal O}+\dots.
    \end{equation}
    
    Next, inspecting the second term and using the cyclicity of the trace and definition of $\mathcal O$, find
    
    \begin{align}
        &\tr{\rho\big[H,[H,\mathcal O]\big]\mathcal O}\\
        &=\tr{H^2\mathcal O^2\rho + H^2 \mathcal O\rho \mathcal O - 2H\mathcal OH\mathcal O\rho}\nonumber\\
        &=2\rho_nO_n\bigg(O_n\langle E_n|H^2|E_n\rangle - \langle E_n|H|E_m\rangle O_m\langle E_m|H|E_n\rangle\bigg).
    \end{align}
    
    The $\mathcal D_P$ states $|E_m\rangle$ are constructed such that they are diagonal in $H$ within rotated subspace $P$, by the TSA procedure. This means that that $\langle E_m|H|E_n\rangle = E_n\delta_{mn}$. However, the operator $H^2$ will generally be different, as it will include states outside of the subspace (see Appendix \ref{app:Eflucs} for details). Generally, $H^2$ may be computed efficiently analytically without needing to use a full Hilbert space. Thus this simplifies to
    
    \begin{align}
    \tr{\rho\big[H,[H,\mathcal O]\big]\mathcal O}&\nonumber\\
    =2\rho_nO_n^2&\bigg(\langle E_n|H^2|E_n\rangle - \big(\langle E_n|H|E_n\rangle\big)^2\bigg).
    \end{align}
    
    The term in parenthesis is the energy variance of eigenstate $|E_n\rangle$ as defined in Eq. \eqref{eq:energyvariance}, and so the decay timescale is related to the energy variance as
    
    \begin{equation}
        \frac{1}{\tau^2} \equiv \frac{\sum_n \rho_nO_n^2\Delta_n^2}{\sum_n \rho_nO_n^2}.
    \end{equation}
    
    \section{AGP Parameters}\label{app:AGPparameters}
    
\begin{figure}[h]
        \centering
        \includegraphics[width=\linewidth]{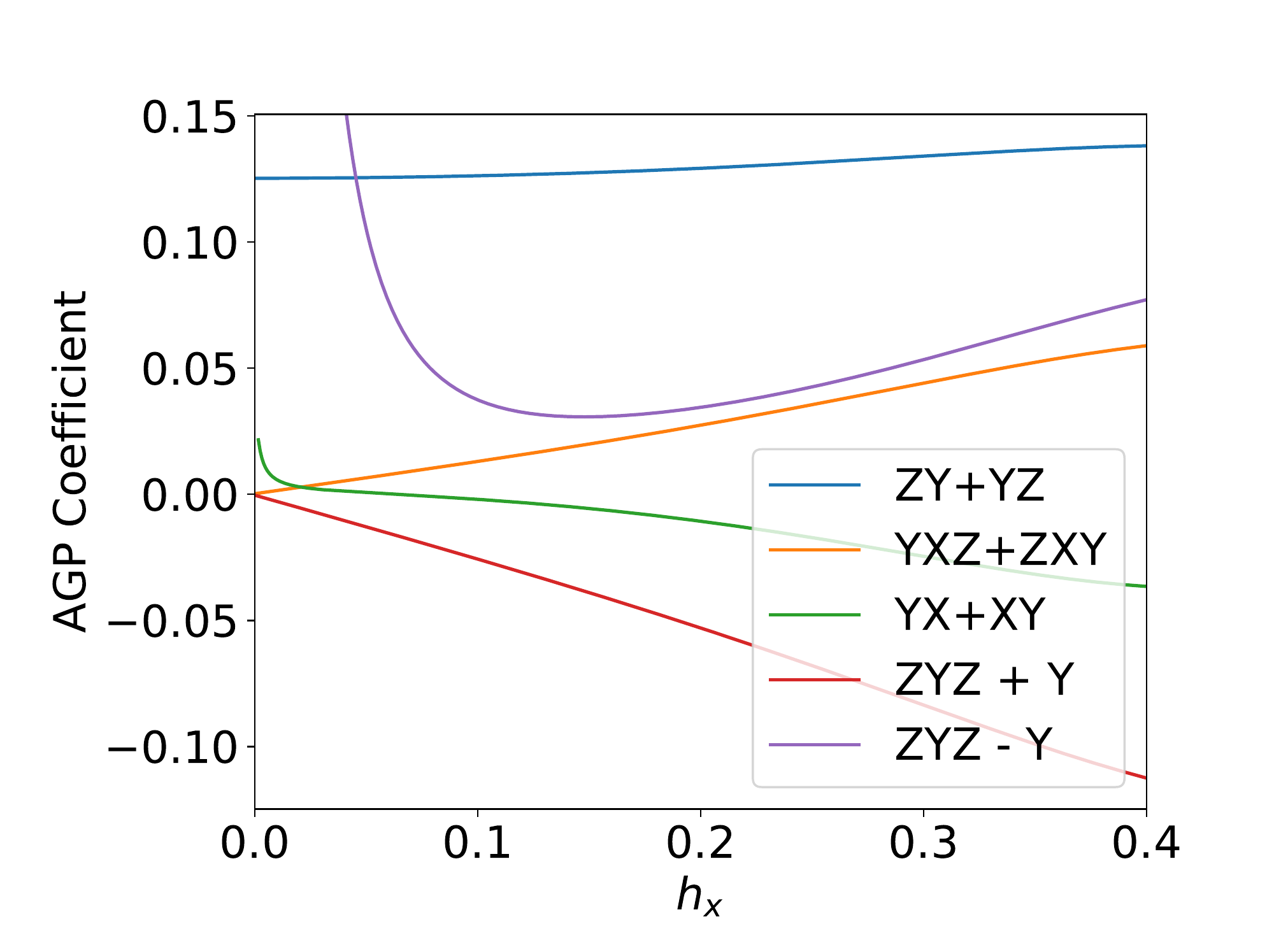}
        \caption{Some of the larger terms in the adiabatic gauge potential along the straight-line path from $(h_x,h_z)=(0,0)\Rightarrow(0.4,0.4)$. Note that one term diverges as $1/h$, but is cutoff numerically when computing the AGP. Notation YX means a translationally invariant sum of Pauli operators on adjacent sites, eg $\sum_i\sigma_y^i\sigma_x^{i+1}$.}
        \label{fig:AGPparams}
    \end{figure}

    For completeness, the coefficients for the variational adiabatic gauge potential are shown in Fig. \ref{fig:AGPparams}. The values are independent of system size and are here computed for 20 sites. Some care needs to be taken with the normalization of the AGP, as it is derived from the differential on some path. Note that some terms diverge close to $h_x=0$, but are cut off numerically.
    
    \normalem
	\bibliographystyle{apsrev4-1}
	\bibliography{citationlist} 
	
\end{document}